\documentclass[doublecol]{epl2}
\usepackage{amssymb}
\usepackage{amsmath}

\title{Low and high intensity velocity selective coherent population trapping in a two-level system}
\author{D. Wilkowski\inst{1} \and M. Chalony\inst{1} \and R. Kaiser\inst{1}  \and A. Kastberg\inst{2}}
\institute{
  \inst{1} Institut Non Lin\'eaire de Nice -
  CNRS, UMR 6618, University de Nice Sophia-Antipolis, F-06560 Valbonne, France\\
  \inst{2} Ume\aa{} University - Department of physics, SE-90187 Ume\aa{}, Sweden
}
\abstract{ An experimental investigation is made of sub-recoil
cooling by velocity selective coherent population trapping in a
two-level system in Sr. The experiment is carried out using the narrow
linewidth intercombination line at 689 nm. Here, the ratio between
the recoil shift and the linewidth is as high as 0.64. We show that,
on top of a broader momentum profile, subrecoil features develop,
whose amplitude is strongly dependent on the detuning from
resonance. We attribute this structure to a velocity selective
coherent population trapping mechanism. We also show that the
population trapping phenomenon leads to complex momentum
profiles in the case of highly saturated transitions, displaying a
multitude of subrecoil features at integer multiples of the recoil
momentum.}

\begin{document}

\maketitle

\section{Introduction}
The role of quantum interference in atomic physic has been
demonstrated in several impressive experiments, \emph{e.g.}, in
\cite{Zibrov:1995,Anderson:1998,Peters:2001}. Already in the
1970'ies, it was shown that quantum interference can prevent
absorption in the presence of resonant light \cite{Alzetta:1976}.
Indeed, if considering a $\Lambda$-shape configuration of internal
states (two ground states and one excited state), there will exist a
coherent superposition of the two ground states, for which the two
excitation amplitudes to the excited state interfere destructively.
For counter propagating laser beams, such a dark state is velocity sensitive, and it can thus
be used for
cooling processes going below the single photon recoil energy.
An experimental proof of such
`velocity selective coherent population trapping' (VSCPT) was first
demonstrated in one dimension in 1988
\cite{Aspect:1988}. The experimental signature consisted of two
peaks in the momentum distribution, centred at the momenta $\pm
\hbar k$, and of sub-recoil widths.
A few years later, VSCPT was also observed in 2D and 3D
\cite{Lawall:1994,Lawall:1995}. Those experiments were done on the
$J_\textrm{g}=1\rightarrow J_\textrm{e}=1$ transition in metastable
helium at 1.08 $\mu$m, using a near resonant laser with
$\sigma^+-\sigma^-$--polarisation configuration. Hence, two major requirements for VSCPT were
fulfilled: firstly, the existence of a closed family of states with
respect to the laser interaction; secondly the resulting dark state
is also an eigenstate of the kinetic energy Hamiltonian. Thus,
the dark state gets an infinitely long lifetime leading to arbitrary
narrow peaks in the momentum distribution. In such conditions, very
low temperatures have been obtained \cite{Saubamea:1997}.

For transitions with higher angular momenta than in the above
mentioned He$^*$-experiments (\emph{i.e.}, with ground state angular
momentum quantum numbers $J_\textrm{g} \geq 2$), at least one dark state exists if
$(J_\textrm{e}=J_\textrm{g})$ or $(J_\textrm{e}=J_\textrm{g}-1)$.
However those dark states are not eigenstates of the
kinetic Hamiltonian. In reference \cite{Olshanii:1992}, the authors
suggest to add an extra off-resonant laser beam to induce
light shifts in order to exactly compensate the kinetic energy mismatch.
Without any compensation, the dark state acquires a finite lifetime
due to motional coupling. As far as the lifetime of this state
remains long with respect to other states, it can be favoured by the
system \cite{Prudnikov:2003}. Thus it may be possible to observe
VSCPT with sub-recoil peaks at momenta $\pm M_J \hbar k$ with
$M_J\geq 1$, where $M_J$ is the magnetic quantum number describing
the projection of $J_\textrm{g}$. So far no experimental data
have been published using such a configuration. High momentum dark states has been reported for the case with a $J_\textrm{g}=1\rightarrow J_\textrm{e}=1$ transition in metastable helium using a lin-angle-lin polarisation configuration \cite{Widmer:1996}. In that case, the dark states are characterised by two peaks at momenta $\pm Q \hbar k$, where $Q$ is an integer. Thus, the high momentum states are still eigenstates of the kinetic Hamiltonian.

From reference \cite{Prudnikov:2003}, we learn that observed momentum states are not necessarily totally dark states, and are thus not eigenstates of the total atom--laser Hamiltonian. However they should be the most protected states with respect to spontaneous emission processes. Following this idea and reference \cite{Widmer:1996}, the requirement of a closed family for VSCPT does not have to be strictly fulfilled. Indeed, considering a two-level atom coupled with a nearly resonant laser, the coherent superposition of ground states with different momenta;
\begin{equation}
\label{dark_state}
\frac{1}{\sqrt{2}} \left( | \textrm{g} \, ; -\hbar k \rangle - | \textrm{g} \, ; +\hbar k \rangle \right)
\end{equation}
is not coupled to the excited state $\left | \textrm{e} \, ; 0
\right \rangle$. However the three involved states, do not form a
closed family because of the coupling to the states $\left |
\textrm{e} \, ; \pm 2\hbar k \right \rangle$. With a broad
transition, \emph{i.e.}, $\varepsilon = \omega_\textrm{r}/ \Gamma
\ll 1$ (where $\Gamma$ is the natural linewidth of the transition
and $\omega_\textrm{r}=\hbar k^2/2m$ is the recoil angular
frequency), the coupling is not sensitive to a kinetic energy
mismatch between the $\left | \textrm{e} \, ; 0 \right \rangle$ and
$\left | \textrm{e} \,  ; \pm 2\hbar k \right \rangle$ states and
VSCPT will not occur. If, however $\varepsilon \approx 1$ (or even
$\varepsilon \gg 1$), the unwanted transitions may be off-resonant,
and the coherent superposition in equation \ref{dark_state} acquire a long lifetime. This situation, leading to VSCPT, has been
numerically studied in \cite{Doery:1995} and was observed on a
helium beam, with $\varepsilon = 0.22$ \cite{Hack:2000}.

In this letter we report on observed VSCPT on the
$^{1}\textrm{S}_{0} \, \rightarrow \, ^{3}\textrm{P}_{1}$
intercombination line in Sr at 689 nm. Here $\varepsilon = 0.64$
(with $\Gamma = 2\pi \, \times \, $7.5 kHz). This more
favorable value than the one used in \cite{Hack:2000} allows for a
more pronounced observation of VSCPT in a two-level system at low intensity.
The momentum distribution is still dominated by Doppler cooling
effects, but an enhanced population builds up at the momenta $\pm
\hbar k$. The amplitudes of these peaks strongly depend on the atom-laser
detuning. Moreover, for a higher saturation parameter, multiple
peaks in the momentum distribution are observed. This structure is
not linked to any closed family of states and, in contrast to the
low saturation case, does not correspond to an eigenstate of the
kinetic Hamiltonian.

\section{Experimental set-up}
The details of the experimental set-up, including two-stage cooling
and trapping of Sr, can be found in
\cite{Chaneliere:2005,Chaneliere:2008}. Briefly, the strontium atoms
are first accumulated and cooled in a (`blue') magneto-optical trap
(MOT) on the $\left( ^{1}\textrm{S}_{0} \, \rightarrow \,
^{1}\textrm{P}_{1} \right)$ transition at 461 nm, and are then
transferred into another (`red') MOT, running on the $\left(
^{1}\textrm{S}_{0} \, \rightarrow \, ^{3}\textrm{P}_{1} \right)$
intercombination line at 689 nm, and with saturation intensity
$I_\textrm{sat} = 3$ $\mu\textrm{W}/\textrm{cm}^{2}$. The cold
sample, with a 50~$\mu$m rms radius, contains about $10^6$ atoms at
a temperature of 1 $\mu$K. After the MOT phases, the atoms interact
with an intensity balanced 1D standing wave along a horizontal
direction. A 0.1 mT vertical magnetic bias field is applied in order to properly define  the quantisation axis. This allows us to restrict the description of the dynamics to two atomic levels. The polarisation of the counter propagating laser beams is adjusted to be parallel to the magnetic field axis. A typical duration of the 1D optical molasses phase is between 0.5 ms and 2 ms. The interaction time is then always longer than the typical lifetime of the relevant internal states. Hence, the studies of the VSCPT cooling mechanism in this article are made in the steady state regime. However, this may not be the case for particular momentum distributions, where a steady state does not exist even for red frequency detuned laser \cite{Castin:1989}.

The momentum distribution is extracted this
distribution using a time-of-flight (TOF) technique. The typical dark
period of the TOF is 50 ms. Thereafter, a resonant probe at 461 nm
is switched on for 40 $\mu$s, and the fluorescence signal is
collected on an intensified CCD camera. The momentum resolution is
limited by the finite size of the cloud after the 1D optical
molasses phase. If this phase is not too long, the typical rms
radius of the cold cloud is still about 50 $\mu$m, which leads to a
momentum resolution of $\hbar k/6$. Images for background
substraction are taken with the same procedure, but with no atoms in
the blue and the red MOTs.

%%%%%%%%%%%%%%%%%%%%%%%%%%%%%%%%%%%%
\begin{figure}
\onefigure[scale=0.28]{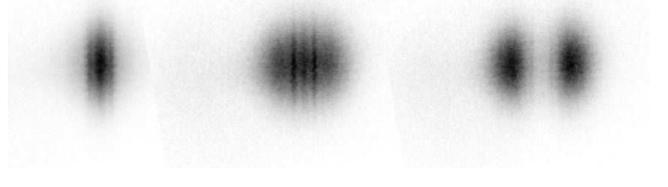}
\caption{Examples of time of flight images. The 1D optical molasses laser beams are along the horizontal
axis in the figure, as well as in the experiment. From left to right, the angular detunings are
$\delta=-3.1 \omega_\textrm{r}$, $-1.8 \omega_\textrm{r}$, and $-0.4 \omega_\textrm{r}$, and the intensities are
$I = 5 I_\textrm{sat}$ for all images, where $I_\textrm{sat}$ is the saturation intensity.} \label{photos}
\end{figure}
%%%%%%%%%%%%%%%%%%%%%%%%%%%%%%%%%%%%

\section{Experimental results}
\begin{figure}
\onefigure[scale=0.22]{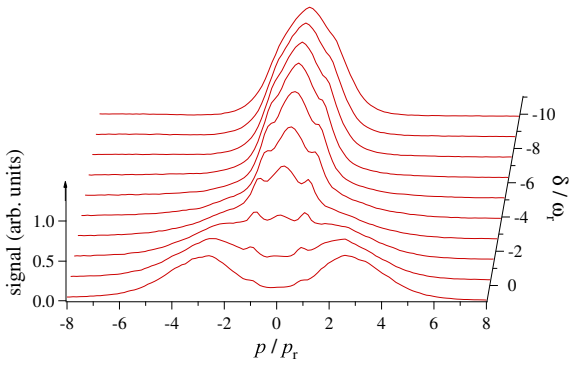} \caption{Momentum distributions for
the intensity $I=5\, I_\textrm{sat}$, and for angular detunings
from $+0.9\, \omega_\textrm{r}$ to $-11.1\, \omega_\textrm{r}$.
These profiles are directly obtained from images as in
fig.~\ref{photos}, by integrating the image density along the
vertical axis, and plotting against the horizontal axis. The latter
is converted to momentum units, taking into account the TOF
expansion time.} \label{waterfall}
\end{figure}

In figure \ref{photos}, we show examples  of acquired TOF-images for
three different detunings, but for otherwise identical laser
parameters. The standing wave axis corresponds to the horizontal
axis of the picture. By integrating along the vertical axis in
fig.~\ref{photos}, we obtain more precise data for the momentum
distributions. In figure \ref{waterfall}, we show such momentum
distributions for a range of detunings.

%%%%%%%%%%%%%%%%%%%%%%%%%%%%%%%%%%%%
%%%%%%%%%%%%%%%%%%%%%%%%%%%%%%%%%%%%

It is known that Doppler cooling on broad transitions ($\varepsilon
\ll 1$) leads to Gaussian momentum distributions.
In the present case, one can clearly see the
more complex distribution resulting from
Doppler cooling on narrow
transitions.
For instance we have observed that
the minimum momentum dispersion is obtained
at a detuning of about
$\delta=-4\omega_\textrm{r}$, in contrast
to the case of broad transitions, where the minimum dispersion occurs at
$\delta =- 0.5\Gamma$, \emph{i.e.}, $\delta \simeq
-0.3\omega_\textrm{r}$ (for $\varepsilon = 0.64$)\cite{Castin:1989}.
For a detuning of $\delta =- 0.5\Gamma$, laser cooling on a narrow
transition yields to two separated maxima, a `double hump': the
atoms are expelled from the central region. This type of
distribution is in qualitative agreement with the predicted
non-stationary distribution reported in \cite{Castin:1989}.

%%%%%%%%%%%%%%%%%%%%%%%%%%%%%%%%%%%%
\begin{figure}
\onefigure[scale=0.4]{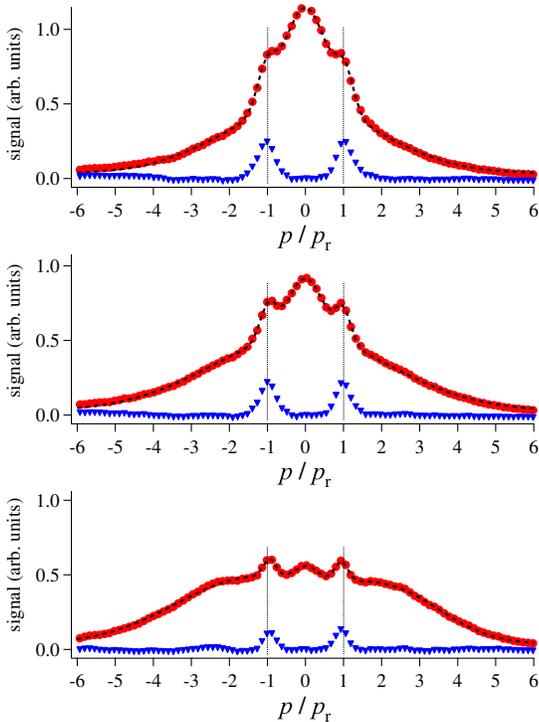} \caption{Momentum profiles from
fig.~\ref{waterfall}, for the angular detunings (from top to bottom),
$-4.4 \omega_\textrm{r}$, $-3.1 \omega_\textrm{r}$, and $-1.8 \omega_\textrm{r}$. The raw
data is the red circles. Fits to four Gaussians (five in the
lowermost case), are presented by a dashed black line. The blue
triangles are the raw data, subtracted by the resulting fit
function, with exception of the two Gaussians that appear close to
$\pm p_\textrm{r}$. Thus, this is a good indication of the part of
the atomic population that is in the semi-dark state. Vertical lines
at $\pm p_\textrm{r}$ are added, as a guide to the eye.}
\label{fit_data}
\end{figure}
%%%%%%%%%%%%%%%%%%%%%%%%%%%%%%%%%%%%

\begin{figure}
\onefigure[scale=0.5]{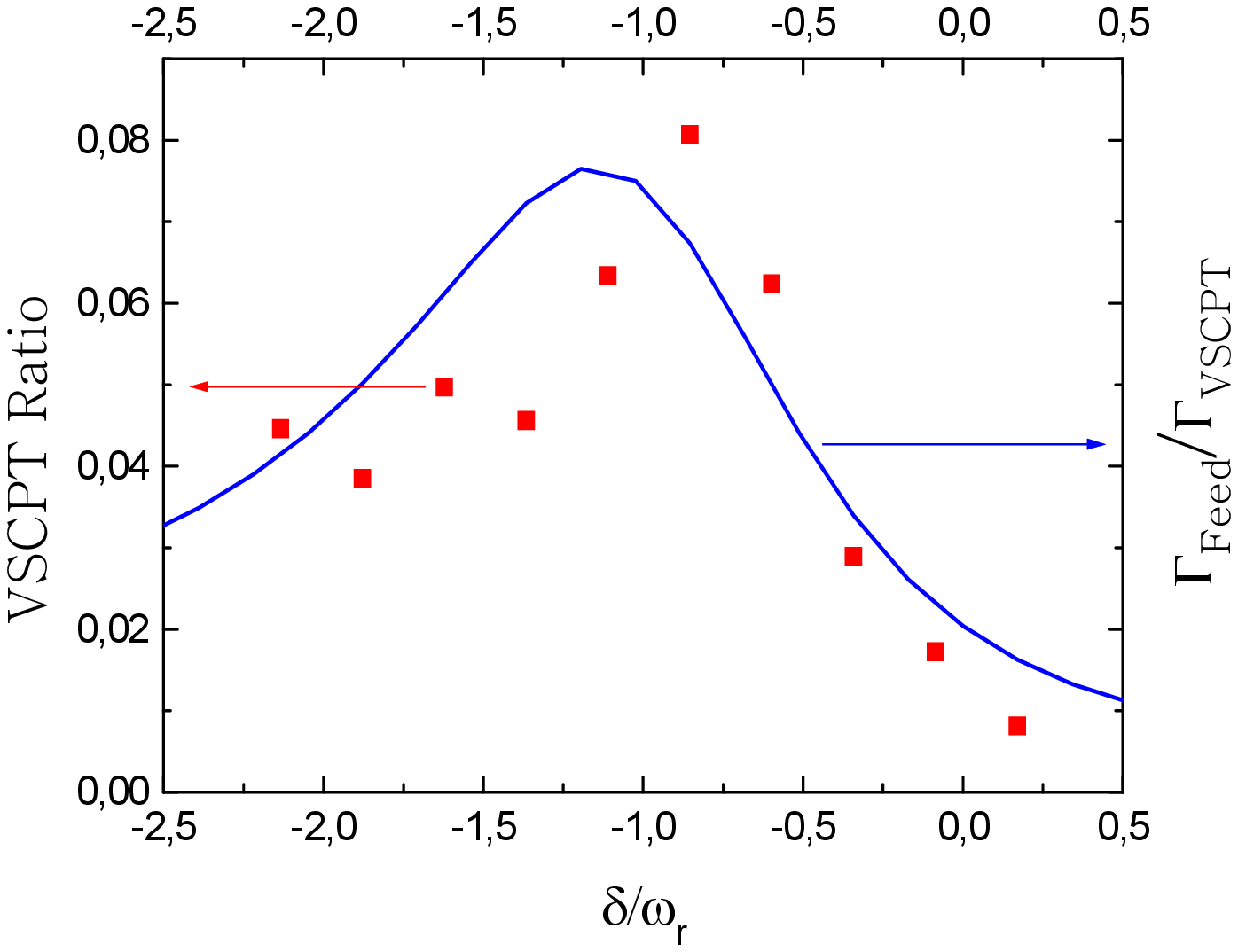} \caption{Red squares: ratio of the
atomic population in the VSCPT state, as a function of angular
detuning. Blue line: ratio of the VSCPT feeding rate
$\Gamma_\textrm{Feed}$ and the VSCPT lifetime
$\Gamma_\textrm{VSCPT}$.} \label{VSCPT-ratio}
\end{figure}

On the overall momentum distribution attributed to Doppler cooling are
superimposed two sub-recoil peaks at $\pm\hbar k$ due to VSCPT \cite{Doery:1995}. The entire momentum distribution is described as a number of Gaussians. Two of those, centered at $p=\pm p_\textrm{r}$, are used to account for VSCPT. In figure \ref{fit_data}, we show such fits, corresponding to three
of the traces in fig.~\ref{waterfall}. The figures also show the residuals from fits including only the broad Doppler features.
The widths of the subrecoil VSCPT features are limited by the convolution with the initial size of the cloud,
and we
can thus not extract relevant information about the VSCPT velocity
distribution. However, by integrating the VSCPT-peaks, we get a measure of the fraction of the population
that is in the long lived (VSCPT) state. In figure~\ref{VSCPT-ratio}
we show the fraction of atoms in the VSCPT-state as a function of
detuning. This quantity remains relatively small, within the few
percent range, and  peaks around $-\omega_\textrm{r}$ below the
atomic resonance.

%%%%%%%%%%%%%%%%%%%%%%%%%%%%%%%%%%%%
%%%%%%%%%%%%%%%%%%%%%%%%%%%%%%%%%%%%

\begin{figure}
\onefigure[scale=0.28]{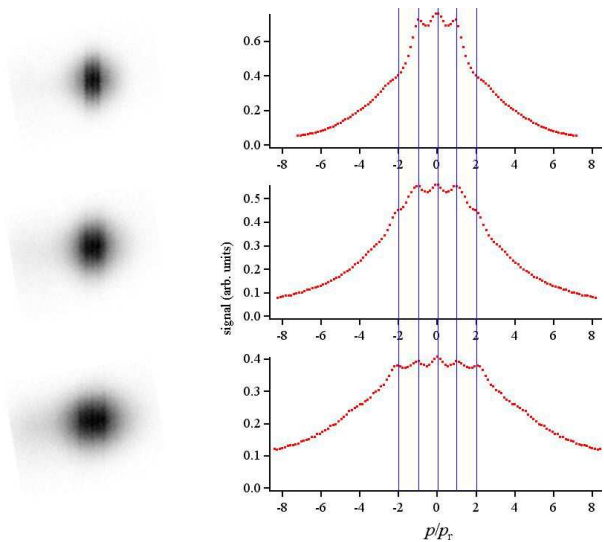} \caption{Time of flight images and
momentum distributions for high intensity. From top to bottom, the
intensities are $I=\, 30 I_\textrm{sat}$, $130\, I_\textrm{sat}$,
and $350\, I_\textrm{sat}$ with $\delta \simeq -5
\omega_\textrm{r}$. In the momentum distribution, vertical lines
have been drawn at integer values of $p/p_\textrm{r}$.}
\label{high_int}
\end{figure}

\begin{figure}
\onefigure[scale=0.9]{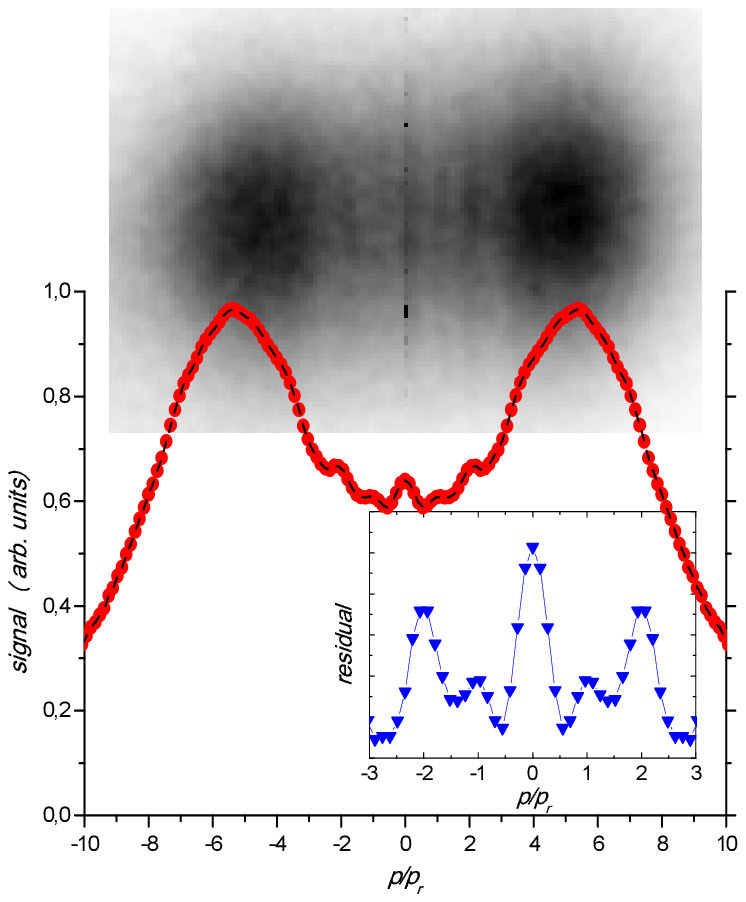} \caption{Time of flight image at
$I=130 I_\textrm{sat}$ and $\delta = -2 \omega_\textrm{r}$, and the
corresponding momentum distribution. To improve the signal-to-noise
ratio the profile is symmetrized with respect to the center of the
distribution. The insert shows a residual, where the slowly varying
Doppler distribution has been removed. } \label{2lambda}
\end{figure}

For higher laser intensity, the momentum distribution becomes more
complex. An increasing number of subrecoil peaks, at momenta $\pm n
\hbar k$ ($n\geq 1$) are now observed, as shown in figure
\ref{high_int}. As we will discuss in the next section, these
subrecoil structures can also be attributed to a VSCPT mechanism.
Those peaks are even less pronounced and broader than for the low
intensity case and a quantitative description is not easy to
achieve. It is however important to note than the population
fractions in the peaks are not necessarily the same or monotonously
decreasing with momentum for a given experimental realisation. For
example figure \ref{2lambda} clearly shows more pronounced peaks for
the $p=0,\pm 2\hbar k$ impulsion than for $p=\pm\hbar k$. We show in
the following section that a long lived state appearing at high
intensity can be attributed to the peaks at $p=0,\pm 2\hbar k$.

%%%%%%%%%%%%%%%%%%%%%%%%%%%%%%%%%%%%
%%%%%%%%%%%%%%%%%%%%%%%%%%%%%%%%%%%%

%%%%%%%%%%%%%%%%%%%%%%%%%%%%%%%%%%%%
%%%%%%%%%%%%%%%%%%%%%%%%%%%%%%%%%%%%

\section{Theoretical model and comparison with experiment}

\subsection{Effective Hamiltonian} A straightforward method to
reveal the existence of long lived states is to diagonalize the
effective Hamiltonian, which takes into account the finite lifetime
of the excited state \cite{Doery:1995}. In the rotating-wave
approximation, it takes the following form:
\begin{eqnarray}
\label{Hgene1}
H_{\textrm{eff}} &=& \frac{p^2}{2m} - \hbar \left( \delta+ i\Gamma\right) | \textrm{e} \, ; p \rangle \langle \textrm{e} \, ; p |  \nonumber\\
&+& \frac{\hbar\Omega}{2} \left( | \textrm{e} \, ; p \rangle \langle \textrm{g} \, ; p+\hbar k | + | \textrm{e} \, ; p \rangle \langle \textrm{g} \, ; p-\hbar k | \right) \, .
\end{eqnarray}
Here $\Omega$ is the Rabi frequency of the atom laser coupling and we have
$\frac{2\Omega^2}{\Gamma^2} = \frac{I}{I_\textrm{sat}}=s_0$, where $I$ is the laser intensity
and $s_0$ the resonant saturation parameter.
The coupling term in the Hamiltonian only connects ground and
excited states that have momentum differences of $\hbar k$. For this
reason, one can reformulate the effective Hamiltonian in the
following way:
\begin{eqnarray}
H_{\textrm{eff}} &=& \sum_{n=-\infty}^{\infty} \left[
\left( \frac {(q+2n\hbar k)^2} {2m} - \hbar \left( \delta+\textrm{i}\Gamma \right)   \right) \right. \nonumber\\
&\times& \left| \textrm{e} \, ; q+n\hbar k \right\rangle \left\langle \textrm{e} \, ; q+n\hbar k \right|  \nonumber\\
&+& \frac{(q+2n\hbar k)^2}{2m} \, \left| \textrm{g} \, ; q+n\hbar k \right\rangle \left\langle \textrm{g} \, ; q+n\hbar k \right|   \nonumber\\
&+& \frac {\hbar\Omega} {2} \left( \, \left|  \textrm{e} \, ; q+n\hbar k \right\rangle \left\langle \textrm{g} \, ; q+(n+1)\hbar k \right| \right. \nonumber\\
&+& \left. \left. \left| \textrm{e} \, ; q+n\hbar k \right\rangle
\left\langle \textrm{g} \, ; q+(n-1)\hbar k \right| \; \right) \;
\right] \, , \label{Hgene2}
\end{eqnarray}
with $n$ being an integer and $0\leq q<\hbar k$. Each
family of states is characterized by a $q$ value and by an odd
(resp. even) value of $n$ for the ground state and an even (resp.
odd) value of $n$ for the excited state.

\begin{figure}
\onefigure[scale=0.52]{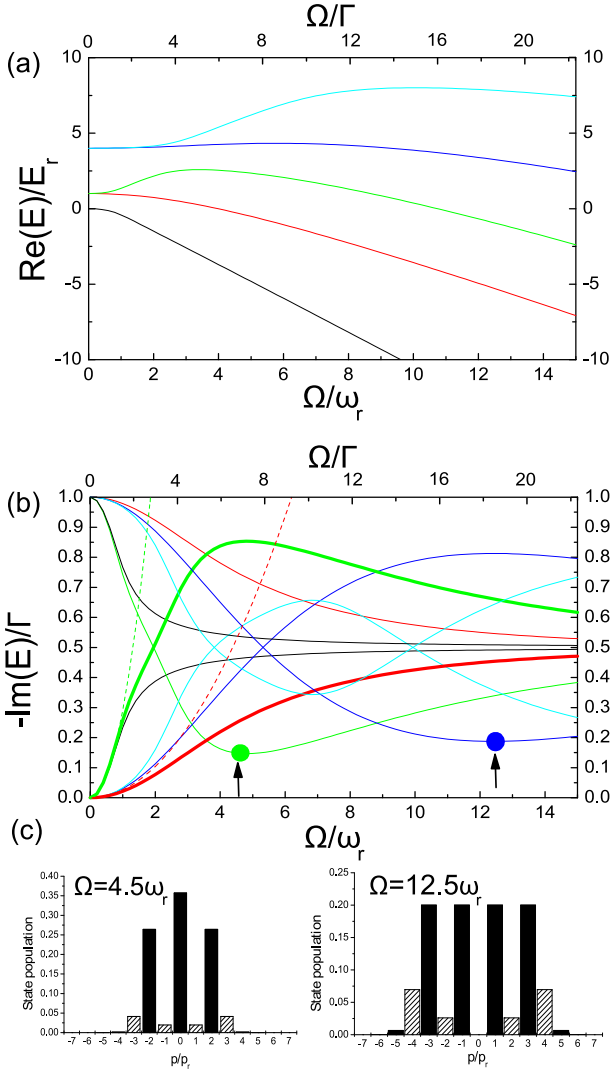} \caption{Real (a) and imaginary
parts (solid lines) (b) of the eigenvalues of the effective
Hamiltonian given by the expression \ref{Hgene2}, with
$n_\textrm{max}=8$ and $q=0$, as a function of $\Omega$. Only the
eigenvalues corresponding to eigenstates for which the momentum
$|p|\leq 2\hbar k$ at $\Omega=0$ are shown. In (b) the dashed red and
black lines correspond respectively to $\Gamma_\textrm{off}$ and
$\Gamma'(\delta+\omega_\textrm{r})$, whereas the dash-dotted line
corresponds to $\Gamma_\textrm{C}$ in units of $\Gamma$ (see text).
At $\Omega \approx 4.5 \omega_\textrm{r}$ and $\Omega \approx 12.5
\omega_\textrm{r}$ long lived states appear, pointed out by arrows
and coloured circles. The histograms in (c) represent the
populations of the relevant momentum states in the ground state
(plain bars) and in the excited state (dashed bars).}
\label{Sim_Num}
\end{figure}

Strictly speaking, each family contains an infinite number of
members. However, one can remove high-momentum states, since the
Doppler shift brings them far off-resonance. Hence to solve the
eigenstate equation, we choose $n_\textrm{max}$, a maximum value of
$n$, such that the coupling between the $\left | \textrm{g}
(\textrm{e}) \, ; q \pm (n_\textrm{max}-1) \hbar k \right\rangle$
state and the $\left | \textrm{e} (\textrm{g}) \, ; q \pm
n_\textrm{max}\hbar k \right \rangle$ state is small and has any impact on the eigenstates of interest. In figs.~\ref{Sim_Num}a
and \ref{Sim_Num}b, we show the real and the imaginary parts of the
eigenvalues, corresponding to some of the eigenstates of the
effective Hamiltonian (equation \ref{Hgene2}) for $\varepsilon = 0.64$ with
$n_\textrm{max}=8$ and $q=0$, as a function of the Rabi frequency.

%%%%%%%%%%%%%%%%%%%%%%%%%%%%%%%%%%%%
%%%%%%%%%%%%%%%%%%%%%%%%%%%%%%%%%%%%

\subsection{Low intensity case} At low Rabi frequency, an expected
behaviour corresponding to VSCPT on a two level system is observed.
%Indeed, one has to consider the $|\textrm{e};0\rangle$ state associate with one dark line and the $1/\sqrt{2}\left( | \textrm{g} \, ; -\hbar k \rangle \pm | \textrm{g} \, ; +\hbar k \rangle \right)$ states associated with one of the red and green lines. For those latest,
The VSCPT state, namely the long lived state, has an eigenvalue
which corresponds to the red line (full line in figure
\ref{Sim_Num}b). Indeed, its imaginary part remains small when
$\Omega$ increases, while its real part does not change
significantly. At vanishing $\Omega$, we check that the eigenstate
is, as expected, the one defined by expression (\ref{dark_state}).
The smooth dependency of the VSCPT state lifetime is mainly due to
the off-resonant coupling to the $|\textrm{e};2\pm\hbar k\rangle$
states. Thus one has
\begin{equation}
\Gamma_\textrm{VSCPT}\xrightarrow[\Omega\rightarrow 0]{}\Gamma_\textrm{off}=\frac{\Gamma\Omega^2}{\Gamma^2+4(\delta-3\omega_\textrm{r})^2},
\end{equation}
where $\Gamma_\textrm{off}$ is the red dashed line in figure \ref{Sim_Num}b. On the other hand, the green
line (in bold in figure \ref{Sim_Num}b) corresponds to the coupling
state orthogonal to the VSCPT state for vanishing $\Omega$. At
resonance, the lifetime $\Gamma_\textrm{C}$ of this state is mainly
due to the coupling to the $|\textrm{e};0\rangle$ state:
\begin{equation}
\Gamma_\textrm{C}\xrightarrow[\Omega\rightarrow 0]{} \Gamma_\textrm{on}=\frac{\Gamma\Omega^2}{\Gamma^2+4(\delta+\omega_\textrm{r})^2}.
\end{equation}
Indeed $\Gamma_\textrm{on}$, which corresponds to the dashed green
line in figure \ref{Sim_Num}b, coincides with the green line at small
$\Omega$.

One can estimate the population of the VSCPT state by a simple feed
and loss mechanism. One then assumes that the population is proportional to the ratio
$\Gamma_\textrm{feed}/\Gamma_\textrm{VSCPT}$, where
$\Gamma_\textrm{feed}\approx(\Gamma_\textrm{VSCPT}+\Gamma_\textrm{C})/2$
is the feeding rate. Using the result given by the numerical
simulation for evaluating
$\Gamma_\textrm{feed}/\Gamma_\textrm{VSCPT}$, we compare this ratio
to the experimental data points in figure \ref{VSCPT-ratio}. The
result is very satisfactory.

\subsection{High intensity case} If the two-level atom can be treated
semi-classically, like for a broad transition for example, the two
imaginary parts monotonously converge at high intensity to the same
value, namely $\textrm{Im}[E]=-\Gamma/2$. In the full quantum
problem, similar behaviours are also observed. This is for example
the case for the states $|e;p=0\rangle$ and $|g;p=0\rangle$ (black
curves in figure \ref{Sim_Num}b). In contrast, other states have
unexpected non-monotonous behaviours. Some of them exhibit minima as
a function of $\Omega$, sort as the one depicted by a \emph{green}
line at $\Omega\approx 4.5 \omega_\textrm{r}$ and a \emph{blue} one
at $\Omega\approx 12.5 \omega_\textrm{r}$. Those states are of
particular interest as they are long lived ones. We will now
demonstrate that those states are at the origin of the multi peak
structures experimentally observed at high intensity.

Figure \ref{Sim_Num}c shows the normalized population of the
momentum states, for the two long lived state. If a two-peak
structure at $p=\pm\hbar k$ is a characteristic of VSCPT at low
intensity, an increasing number of peaks is predicted by our model
and also observed in the experiment at higher intensity. For
example the long lived state at $\Omega\approx 4.5
\omega_\textrm{r}$ is mainly composed of three peaks, at $p=0$ and
at $p=\pm2\hbar k$. At $\Omega\approx 12.5 \omega_\textrm{r}$ a
four-peak structure at $p=\pm\hbar k$ and $p=3\pm\hbar k$ is
expected.

Lets focus on the simplest case, namely the three peak case at
$\Omega\approx 4.5\, \omega_\textrm{r}$, and derive an
analytical expression for the state lifetime. In the view of the
most abundant peak population, we restrict ourselves to
$n_\textrm{max}=2$ and consider only the closed family $\left\{
\left | \textrm{g} \, ; q \right \rangle , \left | \textrm{e} \, ; q
\pm \hbar k \right \rangle ,
 \left | \textrm{g} , q \pm 2\hbar k \right \rangle \right\}$.
Under this condition, a straightforward diagonalisation of the
effective Hamiltonian can be performed in the limit of high
saturation. Here the kinetic terms and the radiative decay are
removed and later been treated with the perturbation theory. One gets the
following eigenvalues; $\lambda=0$, $\lambda= \pm1/2$, and $\lambda=
\pm \sqrt{3}/2$ (in units of $\hbar\Omega$), with the eigenstates
\begin{eqnarray}
\left | \lambda=0 \right \rangle &=&
\frac{1}{\sqrt{3}} \left( \, \left| \textrm{g} \, ; q-2\hbar k \right\rangle -\left| \textrm{g} \, ; q \right \rangle
+ \right. \nonumber \\
&+& \left. \left| \textrm{g} \, ; q+2\hbar k \right\rangle \, \right) \\
\left | \lambda=\pm 1/2 \right \rangle &=&
\frac{1}{2} \left( -\left| \textrm{g} \, ; q-2\hbar k \right\rangle \mp \left| \textrm{e} \, ; q-\hbar k \right\rangle
\pm \right. \nonumber \\
&\pm& \left. \left| \textrm{e} \, ; q+\hbar k \right\rangle + \left| \textrm{g} \, ; q+2\hbar k \right\rangle \, \right) \\
| \lambda=\pm \sqrt{3}/2  \rangle &=&
\frac{1}{\sqrt{12}} \left( \, \left| \textrm{g} \, ; q-2\hbar k \right\rangle
\pm \sqrt{3} \left| \textrm{e} \, ; q-\hbar k \right\rangle \right. + \nonumber \\
&+& 2 \left| \textrm{g} \, ; q \right\rangle \pm \sqrt{3} \left| \textrm{e} \, ; q+\hbar k \right\rangle
+ \nonumber \\
&+& \left. \left| \textrm{g} \, ; q+2\hbar k \right\rangle \, \right) \, .
\end{eqnarray}
In this context the $\left | \lambda=0 \right \rangle$ state has any excited state component with three peaks at $p=0$ and $p=\pm\, 2\hbar k$,
thus very similar to the example shown in fig.~\ref{Sim_Num}c. The
$\left | \lambda=0 \right\rangle$ state gets a finite lifetime
firstly due to the kinetic term which mix it to the $\left |
\lambda=\pm 1/2 \right \rangle$ and $\left | \lambda=\pm 3/2 \right
\rangle$ states and secondly due to the off-resonant coupling to the
$ \left | \textrm{e} \, ; q\pm 3\hbar k \right\rangle$ states. Using
perturbation theory, the imaginary part, of the $\left | \lambda=0
\right\rangle$ state can be calculated. With $q \ll \hbar k$, one
gets:
\begin{equation}
\Gamma_{\lambda=0}=\Gamma_{\textrm{kin}}+\Gamma_{\textrm{off}} \, ,
\label{eq_gamma}
\end{equation}
where
\begin{eqnarray}
\Gamma_{\textrm{kin}} &=& -\frac{(8\omega_\textrm{r})^2}{27}\frac{4\Gamma}{\Omega^2}\left(1+9\left(\frac{q}{\hbar k}\right)^2\right) \nonumber \\
\Gamma_{\textrm{off}} &=& -\frac {\Gamma \Omega^2}{3} \frac{1}
{2(5\omega_\textrm{r})^2+ \Omega^2} \nonumber \\
&\times &\left(1+\frac{(40\omega_r^2)^2}
{(2(5\omega_\textrm{r})^2 + \Omega^2)^2}\left(\frac{q}{\hbar k}\right)^2\right).
\end{eqnarray}
The general dependance of $\Gamma_{\lambda=0}$ at $q=0$, given by
relation \ref{eq_gamma}, is fund to be in good agreement with the
numerical simulation presented in fig.~\ref{Sim_Num}. Moreover, the
$q$ dependence of $\Gamma_{\lambda=0}$ indicates that the long lived
state is also velocity selective. This point is particularly
important for a VSCPT cooling scheme.

Momentum distributions corresponding to the $\left | \lambda=0
\right \rangle$ state have also been observed in the experiment, as
shown in fig.~\ref{2lambda}. The detailed VSCPT distribution is
shown as a residual, where the slow varying envelop has been
numerically removed. We observe three peaks of the same height in
good agreement with the prediction from fig.~\ref{Sim_Num}c. However
the measured intensity is $I \approx 130 I_\textrm{sat}$,
\emph{i.e.}, $\Omega \approx 12 \omega_\textrm{r}$. Even if the
$\left | \lambda=0 \right \rangle$ state is still a long-lived one
at this value, the intensity is two times larger than the predicted
optimum one ($\Omega\approx 4.5 \omega_\textrm{r}$). This
discrepancy may be due to an absence, in our simplified model, of a
dynamical description of the population of the long-lived state.
Indeed the model gives a prediction of the escape rate via the state
lifetime but the feeding process is not described. The observation
of those long lived states may occur at a larger or shifted range of
intensity than the model prediction. Indeed only smooth changes, as
function of the system parameter, on the multi peak structure have
been observed so far. Moroever, since the image of
fig.~\ref{2lambda} does not reveal any peaks at $\pm 3 \hbar k$, the
$\pm \hbar k$ structure can still be associated to the state $\left
( \left | \textrm{g} \, ; -\hbar k \right\rangle - \left |
\textrm{g} \, ; +\hbar k\right \rangle \right) /\sqrt{2}$,
corresponding to the low intensity case.

\section{Conclusions}

VSCPT-cooling on a two-level atomic system has been experimentally
demonstrated. This is possible, since the atomic transition used, in
Sr, has a narrow linewidth, making the otherwise open family of
momentum states, $ \left | \textrm{g} \, ; -\hbar k \right \rangle$,
$\left | \textrm{e} \, ; 0 \right \rangle$, and $\left | \textrm{g}
\, ; +\hbar k \right \rangle$, less open, due to kinetic detuning
from other momentum states. Thus, a semi-dark state is formed.
During the cooling, the long-lived state, of sub-recoil width, is
fed with atoms, while there is simultaneously a constant loss. For a
favourable ratio between the feeding rate and the loss rate, the
momentum profile acquires narrow peaks, centered at $\pm \hbar k$,
that lie on top of the Doppler cooled profile. At best, the
steady-state population of the dark state reaches $\approx 10\%$.

At high saturation, complex momentum profiles arise with subrecoil
peaks also resulting from a coherent population trapping mechanism.
With an analysis based on an effective Hamiltonian approach, we have
identified the observed multipeak structures to long lived states.

\acknowledgments Freddy Bouchet, Claude Dion and Mats Nyl\'en are
kindly acknowledged for discussions. A.K. thanks the
\emph{International Cold Atom Network (INTERCAN)} for support, and
the staff at INLN for hospitality. This work was financially support
by the \emph{Conseil g\'{e}n\'{e}ral des Alpes-Maritimes} and the
\emph{Laboratoire National de M\'{e}trologie et d'Essai (LNE)}.

\end{document}